\documentclass{article}
\usepackage{spconf,amsmath,graphicx}
\usepackage{booktabs}
\usepackage{amssymb}
\usepackage{bm}
\usepackage{multirow}
\usepackage{arydshln}
\usepackage{cite}
\usepackage{url}
\usepackage{setspace}
\usepackage{xcolor,soul}

\definecolor{red}{rgb}{1,0,0}

\makeatletter
\def\adl@drawiv#1#2#3{%
        \hskip.5\tabcolsep
        \xleaders#3{#2.5\@tempdimb #1{1}#2.5\@tempdimb}%
                #2\z@ plus1fil minus1fil\relax
        \hskip.5\tabcolsep}
\newcommand{\cdashlinelr}[1]{%
  \noalign{\vskip\aboverulesep
           \global\let\@dashdrawstore\adl@draw
           \global\let\adl@draw\adl@drawiv}
  \cdashline{#1}
  \noalign{\global\let\adl@draw\@dashdrawstore
           \vskip\belowrulesep}}
\makeatother

\newcommand{\pz}{\phantom{0}}

\newcommand\independent{\protect\mathpalette{\protect\independenT}{\perp}}
\def\independenT#1#2{\mathrel{\rlap{$#1#2$}\mkern2mu{#1#2}}}

\DeclareMathOperator*{\argmax}{argmax}

\usepackage{cleveref}
\crefformat{section}{\S#2#1#3}
\crefformat{subsection}{\S#2#1#3}
\crefformat{subsubsection}{\S#2#1#3}

\renewcommand{\paragraph}[1]{\noindent\textbf{#1}\hspace{0.5em}}

\title{InterMPL: Momentum Pseudo-Labeling with Intermediate CTC Loss}

\name{
    Yosuke Higuchi$^{1,2}$,
    Tetsuji Ogawa$^{2}$,
    Tetsunori Kobayashi$^{2}$,
    Shinji Watanabe$^{1}$
}

\address{
    $^1$Carnegie Mellon University, USA\ \ $^2$Waseda University, Japan
}

\begin{document}
\ninept
\maketitle
\setlength{\abovedisplayskip}{4pt}
\setlength{\belowdisplayskip}{4pt}
\setlength{\textfloatsep}{0.4cm} %
\begin{abstract}
This paper presents InterMPL,
a semi-supervised learning method of end-to-end automatic speech recognition (ASR)
that performs pseudo-labeling (PL) with intermediate supervision.
Momentum PL (MPL) trains a connectionist temporal classification (CTC)-based model on unlabeled data
by continuously generating pseudo-labels on the fly and improving their quality.
In contrast to autoregressive formulations,
such as the attention-based encoder-decoder and transducer,
CTC is well suited for MPL,
or PL-based semi-supervised ASR in general,
owing to its simple/fast inference algorithm and
robustness against generating collapsed labels.
However, CTC generally yields inferior performance than the autoregressive models due to the conditional independence assumption,
thereby limiting the performance of MPL.
We propose to enhance MPL by introducing intermediate loss,
inspired by the recent advances in CTC-based modeling.
Specifically,
we focus on self-conditional and hierarchical conditional CTC,
that apply auxiliary CTC losses to intermediate layers
such that the conditional independence assumption is explicitly relaxed.
We also explore how pseudo-labels should be generated and used as supervision for intermediate losses.
Experimental results in different semi-supervised settings demonstrate that the proposed approach outperforms MPL and improves an ASR model by up to a 12.1\% absolute performance gain.
In addition, our detailed analysis validates the importance of the intermediate loss.
\end{abstract}

\vspace{-0.2cm}
\section{Introduction}
\label{sec:intro}
\vspace{-0.2cm}
End-to-end (E2E) automatic speech recognition (ASR)~\cite{graves2014towards,chorowski2015attention,chan2016listen} has achieved remarkable improvements in performance
thanks to innovative sequence-to-sequence modeling techniques~\cite{graves2006connectionist,graves2012sequence,sutskever2014sequence,bahdanau2014neural} with
sophisticated neural network architectures~\cite{dong2018speech,kriman2020quartznet,gulati2020conformer}.
While E2E ASR has shown promising results on a variety of benchmarks~\cite{chiu2018state, luscher2019rwth, karita2019comparative},
training E2E ASR is generally data-hungry:
its performance often relies on the availability of abundant labeled (transcribed) speech data~\cite{kahn2020libri},
which is not always achievable due to high annotation costs.

In order to mitigate the heavy requirement on labeled data,
semi-supervised learning has been actively studied in E2E ASR,
utilizing a large quantity of unlabeled speech-only data to enhance the model performance.
Among diverse semi-supervised learning approaches,
pseudo-labeling (PL)~\cite{lee2013pseudo} has been gathering attention
due to its simple yet effective training algorithm~\cite{li2019semi,masumura2020sequence,weninger2020semi,hsu2020semi,xu2020iterative,chen2020semi,park2020improved,likhomanenko2021slimipl,moritz2021semi}.
In typical PL,
a teacher (seed) model is first trained on labeled data
and used to generate pseudo-labels by transcribing unlabeled data.
A student model is then trained using the labeled and pseudo-labeled data,
with the aim of performing better than the teacher.
Shallow fusion is often performed during the labeling process,
which utilizes an external language model (LM) to generate higher-quality pseudo-labels~\cite{kahn2020self, hsu2020semi}.
Data augmentation also plays an important role in providing the student with informative training signals~\cite{chen2020semi,masumura2020sequence,weninger2020semi}.
In addition to the techniques above,
iterating the PL process has shown to be essential for improving ASR performance,
periodically updating pseudo-labels as the model training proceeds~\cite{chen2020semi,xu2020iterative,park2020improved,likhomanenko2021slimipl}.
Momentum PL (MPL)~\cite{higuchi2021momentum,manohar2021kaizen,higuchi2022advancing,higuchi2022momentum} is one of the recent iterative methods,
inspired by the mean teacher framework~\cite{tarvainen2017mean}.
MPL trains a pair of offline ($\approx$ teacher) and online ($\approx$ student) models
that interact and learn from each other.
In each training step,
pseudo-labels are generated on the fly by the offline model via greedy decoding
and used as targets to train the online model.
The offline model maintains an exponential moving average of the online model weights to stabilize the label generation.
Through the interaction between the two models,
MPL enables continuous updates on pseudo-labels and,
concurrently, improves the label quality.

Connectionist temporal classification (CTC)~\cite{graves2006connectionist} is a promising approach for conducting iterative PL, particularly MPL,
compared to other autoregressive models equipped with a recurrent decoder (e.g., attention-based encoder-decoder~\cite{chorowski2015attention,chan2016listen} and transducer~\cite{graves2012sequence}).
The non-autoregressive formulation in CTC allows a model to transcribe unlabeled speech data efficiently with its simple, fast and parallelized inference algorithm,
a crucial property for the on-the-fly label generation in MPL.
Furthermore,
CTC is robust against generating collapsed pseudo-labels,
which is frequently caused by autoregressive decoding (e.g., word skipping or repeating~\cite{chorowski2016towards,kahn2020self}).
This does not necessarily require a model to apply heuristic filtering techniques for excluding erroneous labels that hinder semi-supervised training~\cite{park2020improved}.
However, the ASR performance of CTC often lags behind those of the autoregressive models~\cite{chiu2018state}.
This is attributed to the fact that CTC assumes output tokens are conditionally independent of each other,
making a model less capable of capturing contextual information.

Hence, our work aims to further enhance the MPL performance
while maintaining the advantages of CTC-based modeling.
To this end,
we propose to introduce intermediate CTC loss~\cite{tjandra2020deja,lee2021intermediate} to MPL,
given the recent advances in non-autoregressive E2E ASR~\cite{higuchi2021comparative}.
We consider adopting
self-conditional CTC (SC-CTC)~\cite{nozaki2021relaxing} and hierarchical conditional CTC (HC-CTC)~\cite{higuchi2022hierarchical}
for MPL.
SC-CTC applies auxiliary CTC losses to intermediate model layers and
utilizes each intermediate prediction as a condition for subsequent layers.
This induces the contextualization of representations,
which is beneficial for relaxing the conditional independence assumption.
HC-CTC extends SC-CTC by
gradually increasing the granularity of each output sequence
in a hierarchical manner~\cite{sanabria2018hierarchical,krishna2018hierarchical},
where the hierarchical structure allows a model to learn the progressive generation of a target sequence.
Through SC-CTC and HC-CTC,
a model is capable of generating multiple pseudo-labels from its intermediate layers.
We thereby explore how pseudo-labels should be generated and used as supervision for intermediate losses during MPL training.

\begin{figure*}[t]
    \begin{minipage}[b]{0.33\linewidth}
        \centering
        \includegraphics[height=3.8cm]{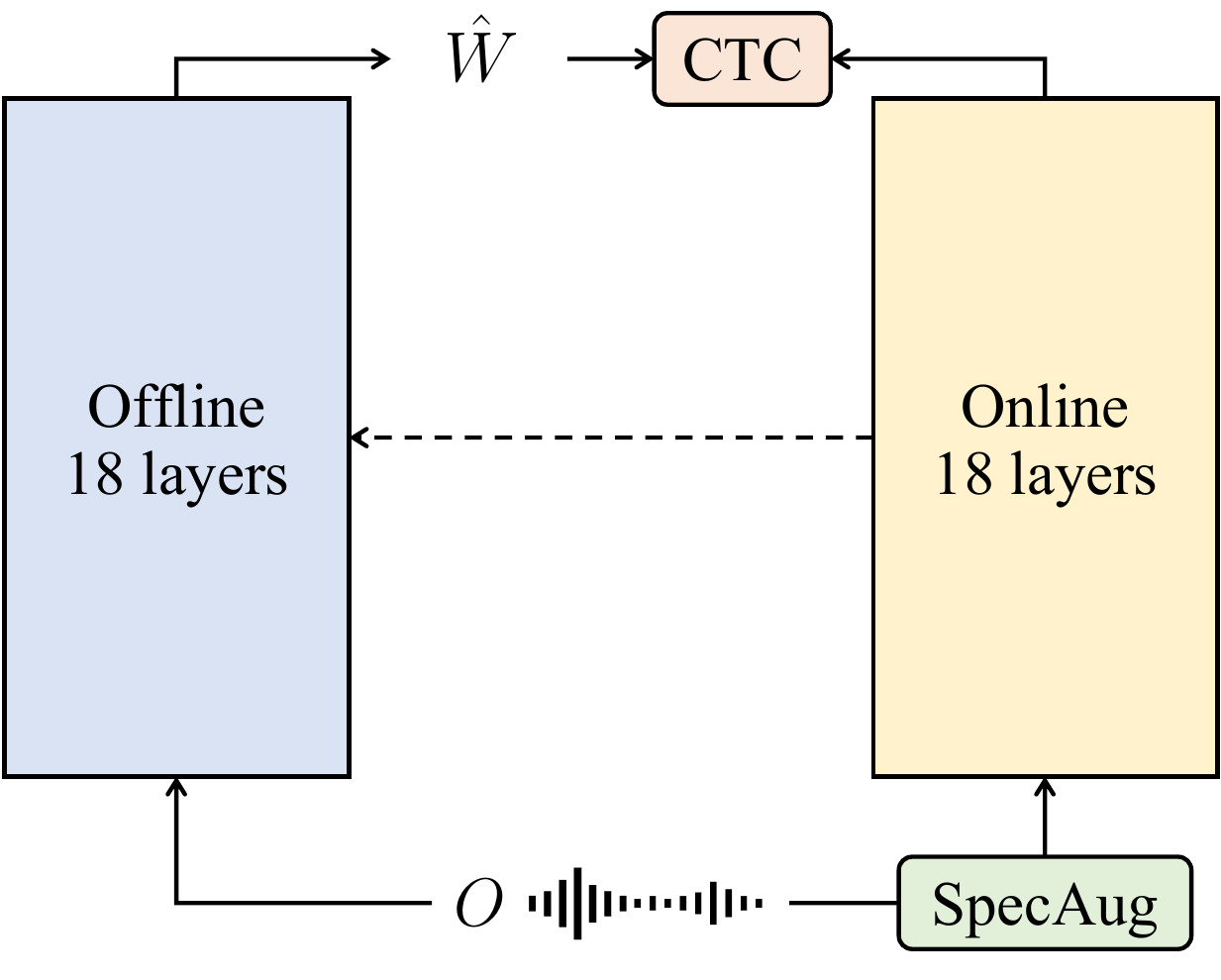}\\
        (a) MPL
    \end{minipage}
    \begin{minipage}[b]{0.33\linewidth}
        \centering
        \includegraphics[height=3.8cm]{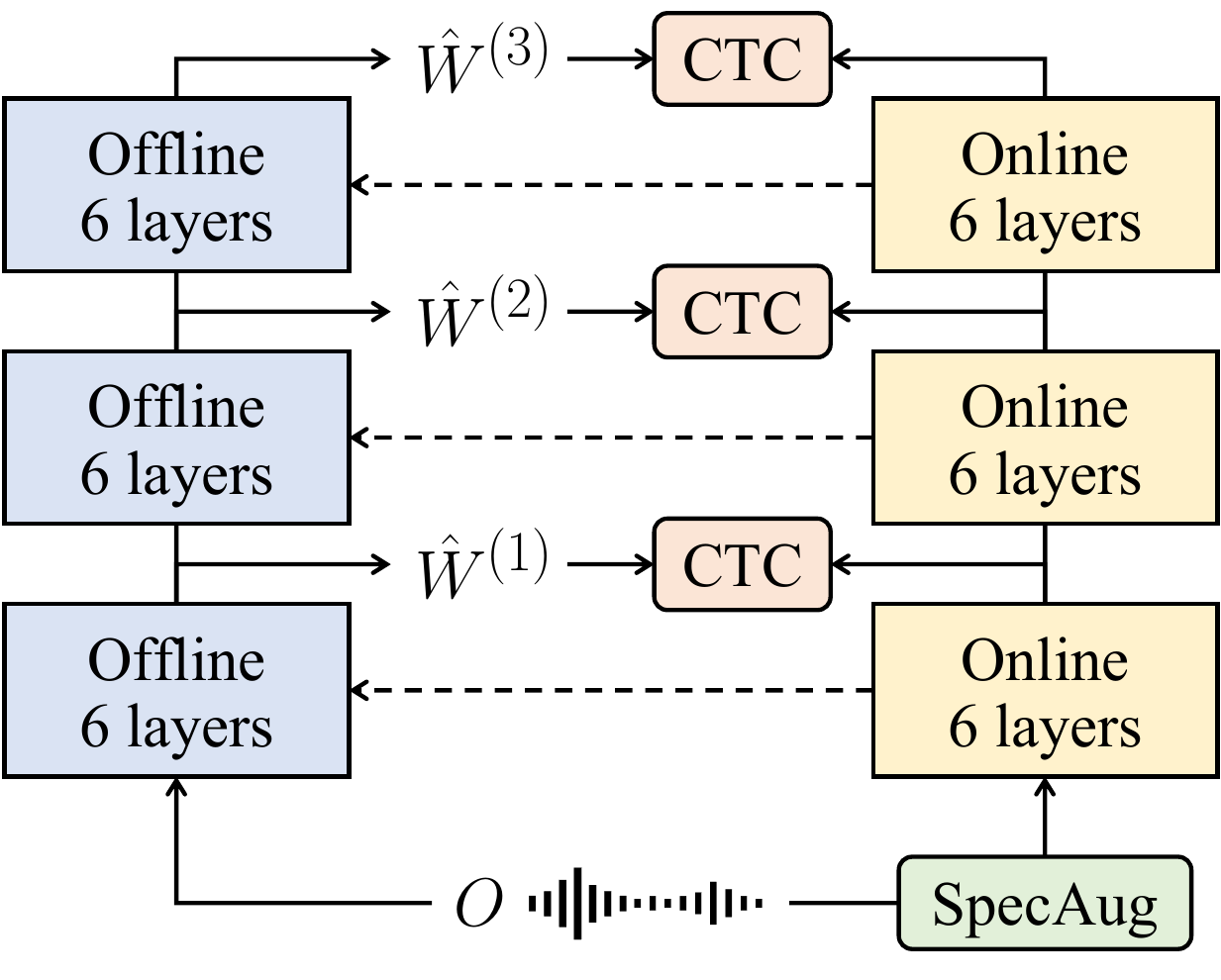}\\
        (b) InterMPL
    \end{minipage}
    \begin{minipage}[b]{0.33\linewidth}
        \centering
        \includegraphics[height=3.8cm]{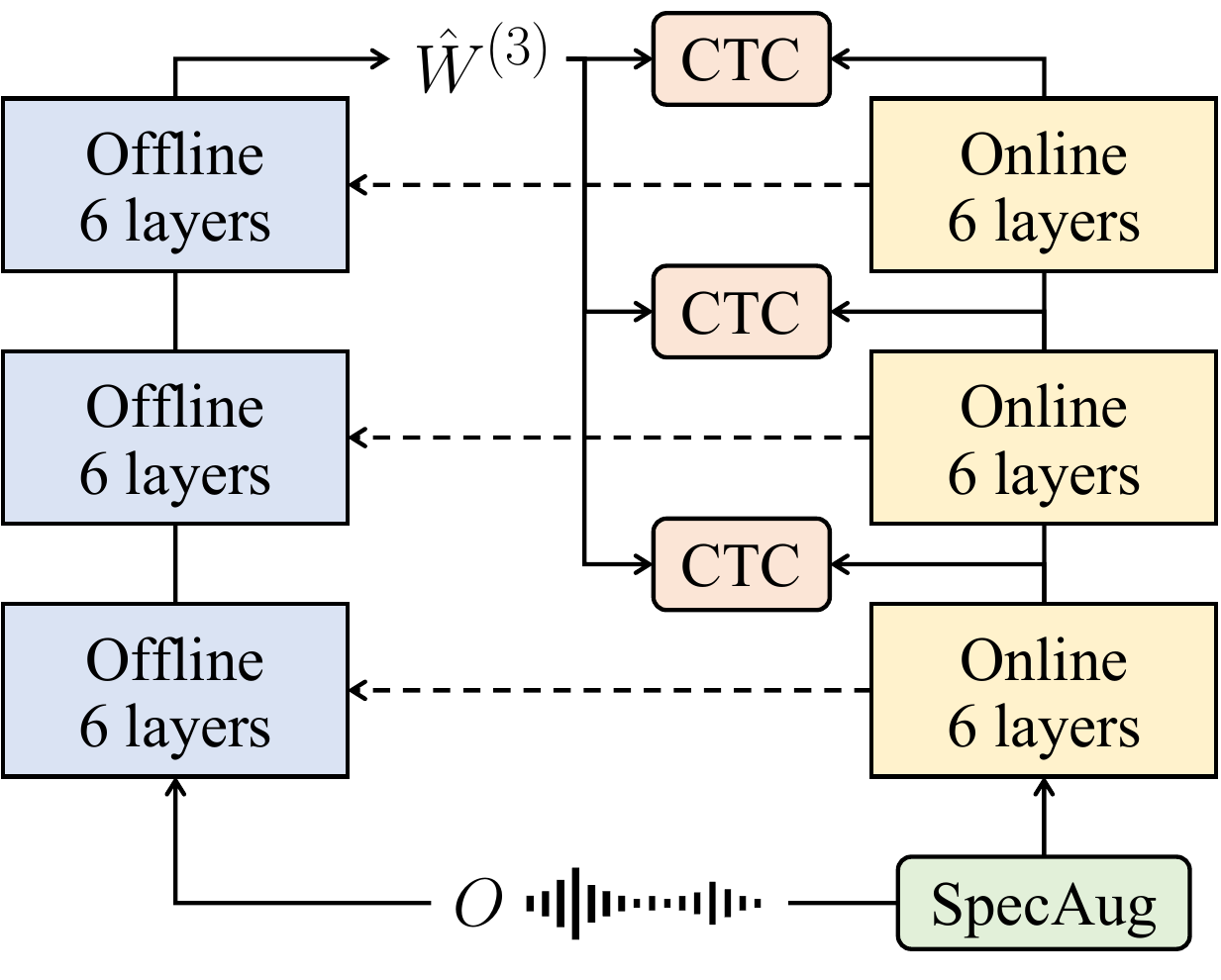}\\
        (c) InterMPL-Last
    \end{minipage}
    \vspace{-0.6cm}
    \caption{Comparisons between conventional MPL and proposed InterMPL for semi-supervised ASR.
    A dashed line ($\dashleftarrow$) indicates the momentum update of the offline model using the online model parameters. The number of total losses is set to three for InterMPL (i.e., $|\mathcal{K}|=3$).}
    \vspace{-0.35cm}
    \label{fig:intermpl}
\end{figure*}

The key contributions of our work are summarized as follows:
1) We propose InterMPL, which enhances MPL-based semi-supervised ASR by intermediate CTC loss;
2) Experimental results show that InterMPL significantly outperforms MPL in various semi-supervised scenarios. We also present a detailed analysis to validate the significance of the intermediate loss; and
3) The codes and recipes are made publicly available at {\footnotesize \url{https://github.com/YosukeHiguchi/espnet/tree/intermpl}}.

\vspace{-0.25cm}
\section{Background}
\vspace{-0.2cm}
\subsection{CTC-based End-to-End ASR with Intermediate Loss}
\vspace{-0.15cm}
E2E ASR is defined as a sequence-mapping problem between a $T$-length input sequence $O \!=\! (\bm{\mathrm{o}}_t \in \mathbb{R}^F| t\!=\!1,\dots,T)$ and 
$U$-length output sequence $W \!=\! ( w_u \in \mathcal{V} | u\!=\!1,\dots,U)$,
where $\bm{\mathrm{o}}_t$ is an $F$-dimensional acoustic feature at frame $t$, $w_u$ is an output token at position $u$, and $\mathcal{V}$ is a vocabulary.
For embedding $O$ into a latent representation space,
we construct a Conformer-based model~\cite{gulati2020conformer}
consisting of $K$ encoder layers.
The $k$-th layer takes as input a previous sequence $H^{(k-1)}$ and outputs a sequence $H^{(k)}$ as
\begin{equation}
    \label{eq:encoder}
    H^{(k)} = \text{Encoder}^{(k)}(H^{(k-1)}),
\end{equation}
where $H^{(0)}\!=\!O$, and $H^{(k)}\!=\!(\bm{\mathrm{h}}_t^{(k)}\!\in\!\mathbb{R}^{D} | t\!=\!1,\cdots,T)$ is a sequence of $D$-dimensional hidden vectors with the same length as that of $O$.
For simplicity, we define $H$ as the last sequence $H^{(K)}$.

\paragraph{CTC}
CTC~\cite{graves2014towards} formulates E2E ASR by considering all possible alignments between $O$ and $W$.
To align the sequences at the input frame level,
CTC augments an output sequence by allowing consecutive identical tokens and inserting a blank symbol $\epsilon$.
Let $A\!=\!(a_t\!\in\!\mathcal{V} \cup \{\epsilon\} | t\!=\!1,\cdots,T)$ be an augmented output sequence, which we refer to an alignment path between $O$ and $W$.
CTC trains a model to predict the paths by minimizing the following loss:
\begin{align}
    \label{eq:L_ctc}
    \mathcal{L}_{\mathsf{ctc}}(W | H) = - \log \sum_{A \in \mathcal{B}^{-1} (W)} \prod_{t} p(a_t | H),
\end{align}
where $\mathcal{B}(\cdot)$ is the collapsing function~\cite{graves2006connectionist} that maps $A$ to $W$ by suppressing repeated tokens and removing blank symbols, and $\mathcal{B}^{-1}(W)$ is a set of all possible paths compatible with $W$.

\paragraph{Intermediate CTC with conditioning} Intermediate CTC~\cite{lee2021intermediate} applies
auxiliary CTC losses to the intermediate hidden-state of the encoder.
In addition to the original CTC loss applied to the last layer (Eq.~\eqref{eq:L_ctc}),
the intermediate losses are calculated as
\begin{equation}
    \label{eq:L_ic}
    \mathcal{L}_{\mathsf{ic}}(W | O) = \sum_{k \in \mathcal{K}} \mathcal{L}_{\mathsf{ctc}}(W | H^{(k)}),
\end{equation}
where $\mathcal{K}$ is a set of layer indices where the CTC losses are computed, and
we equally distribute the weight across the losses.
Note that Eq.~\eqref{eq:L_ic} always includes the last loss (i.e., $K\!\in\!\mathcal{K}$) and
is equal to Eq.~\eqref{eq:L_ctc} when $\mathcal{K}\!=\!\{K\}$.
Self-conditional CTC (SC-CTC)~\cite{nozaki2021relaxing} extends Intermediate CTC by conditioning the encoder using a sequence predicted from each intermediate layer.
Specifically, after calculating $H^{(k)}$ from Eq.~\eqref{eq:encoder} at an intermediate layer,
SC-CTC adds a sequence of posterior probability distributions $P^{(k)}\!=\!(p(a_t|H^{(k)})\!\in\![0,1]^{|\mathcal{V}|+1}|t\!=\!1,\cdots,T)$ as
\begin{equation}
    H^{(k)} \leftarrow H^{(k)} + \text{Linear}_{|\mathcal{V}|+1\rightarrow D} (P^{(k)}),
    \label{eq:cond}
\end{equation}
where $k\!\in\!\mathcal{K}$.
This has been shown to further improve Intermediate CTC by relaxing the conditional independence assumption in Eq.~\eqref{eq:L_ctc} ($a_t\!\independent\!a_{\ne t} | H$).
Hierarchical-conditional CTC (HC-CTC)~\cite{higuchi2022hierarchical} extends SC-CTC by hierarchically increasing the output unit size of each intermediate prediction, using different subword vocabularies.

\vspace{-0.2cm}
\subsection{Semi-Supervised ASR with Momentum Pseudo-Labeling}
\vspace{-0.15cm}
\label{ssec:mpl}
In semi-supervised ASR, a seed model is first trained on labeled data $\mathcal{D}_{\mathsf{lab}}\!=\!\{ \langle O_n, W_n \rangle | n\!=\!1,\dots,N \}$ using the CTC loss from Eq.~\eqref{eq:L_ctc}.
Momentum pseudo-labeling (MPL)~\cite{higuchi2021momentum} is then applied to the seed model to improve the performance using
unlabeled speech-only data $\mathcal{D}_{\mathsf{unlab}} \!=\! \{ O_m | m\!=\!N\!+\!1,\dots,N\!+\!M \}$.
Figure~\ref{fig:intermpl}(a) illustrates the training process of MPL based on a pair of \textit{online} and \textit{offline} models. Let $\xi$ and $\phi$ denote the parameters of the online and offline models,
which are initialized with the pre-trained seed model parameters.

\paragraph{Online model training}
Given the $m$-th unlabeled sample $O_m\!\in\!\mathcal{D}_{\mathsf{unlab}}$ and its encoded sequence $H_m$,
the online model is trained on pseudo-labels $\hat{W}_m$ generated on the fly by the offline model with $\phi$:
\begin{equation}
    \label{eq:W_hat}
    \hat{W}_m = \mathcal{B} (\argmax_{a_t} p (a_t | H_m, \phi) | t=0,\cdots,T).
\end{equation}
With the pseudo-labeled sample $\langle O_m, \hat{W}_m \rangle$,
the online model with $\xi$ is trained via a gradient descent optimization based on the CTC loss $\mathcal{L}_{\mathsf{ctc}}(\hat{W}_m|H_m,\xi)$ from Eq.~\eqref{eq:L_ctc}.
Here, the input speech is augmented by SpecAugment~\cite{park2019specaugment} (as shown in Fig.~\ref{fig:intermpl}(a)) to facilitate the model training on pseudo-labels.
Note that MPL also uses the $n$-th labeled sample $\langle O_n, W_n \rangle\!\in\!\mathcal{D}_{\mathsf{lab}}$ and trains the online model
with supervised loss $\mathcal{L}_{\mathsf{ctc}}(W_n|H_n,\xi)$,
which helps the online model stabilize and promote learning from unlabeled data.

\paragraph{Offline model training}
After every update of the online model,
the offline model accumulates the parameters of the online model as
$\phi \leftarrow \alpha \phi + (1 - \alpha) \xi$,
an exponential moving average with a momentum coefficient $\alpha\!\in\!(0, 1)$.
This momentum update makes
the offline model serve as an ensemble of the online models at different training steps,
stabilizing and reinforcing the label generation in Eq.~\eqref{eq:W_hat}.

Through the above interaction between the two models,
MPL realizes stable and continuous ASR training on unlabeled data,
concurrently improving the quality of pseudo-labels.

\vspace{-0.25cm}
\section{InterMPL}
\vspace{-0.2cm}
We propose a semi-supervised ASR method that introduces the intermediate CTC loss to MPL.
The conventional MPL is founded on the CTC-based modeling,
whose performance can be limited due to the conditional independence assumption (cf.\ Eq.~\eqref{eq:L_ctc}).
To further enhance MPL,
we adopt SC-CTC or HC-CTC for constructing a seed model,
which is expected to facilitate better CTC training/decoding and thus promote the succeeding semi-supervised process with higher-quality pseudo-labels.
Given labeled data $\mathcal{D}_{\mathsf{lab}}$,
a seed model is trained by a supervised loss $\mathcal{L}_{\mathsf{ic}}(W_n|O_n)$ from Eq.~\eqref{eq:L_ic} along with the conditioning mechanism in Eq.~\eqref{eq:cond}.

Initialized with the seed model trained by SC-CTC or HC-CTC,
the online model can accept intermediate supervision using pseudo-labels, and
the offline model can generate multiple pseudo-labels from its intermediate layers.
This motivated us to consider two approaches,
namely \textbf{InterMPL} (Fig.~\ref{fig:intermpl}(b)) and \textbf{InterMPL-Last} (Fig.~\ref{fig:intermpl}(c)),
for fully utilizing the intermediate mechanism in MPL.

\paragraph{InterMPL}
In InterMPL, the offline model generates pseudo-labels from each prediction layer,
as shown in Fig.~\ref{fig:intermpl}(b) with three different outputs.
These pseudo-labels are used to calculate a loss for the corresponding layer of the online model.
Given the $m$-th unlabeled sample $O_m\!\in\!\mathcal{D}_{\mathsf{unlab}}$,
the $k$-th offline encoder layer emits hidden vectors $H_m^{(k)}$ and
generates a $k$-th prediction $\hat{W}_m^{(k)}$ as in Eq.~\eqref{eq:W_hat} as
\begin{equation}
    \label{eq:W_hat_k}
    \hat{W}_m^{(k)} = \mathcal{B} (\argmax_{a_t} p (a_t | H_m^{(k)}, \phi) | t=0,\cdots,T),
\end{equation}
where $k\!\in\!\mathcal{K}$.
With the unlabeled input and multiple pseudo-labels $\langle O_m, \{\hat{W}_m^{(k)}\}_{k\in\mathcal{K}} \rangle$,
the objective function of the online model is defined based on Eq.~\eqref{eq:L_ic} as
\begin{equation}
    \mathcal{L}_{\mathsf{ic}}(\{\hat{W}_m^{(k)}\}_{k\in\mathcal{K}} |O_m,\xi) = \sum_{k \in \mathcal{K}} \mathcal{L}_{\mathsf{ctc}}(\hat{W}_m^{(k)} | H^{(k)}, \xi).
\end{equation}
This training strategy is compatible with both SC-CTC and HC-CTC-based InterMPL,
which we assume particularly effective for HC-CTC with varying output units.
HC-CTC trains an ASR model to learn a progressive generation of a target sequence,
using the intermediate loss with increasing subword vocabulary size.
We expect pseudo-labels generated at different granularities to facilitate semi-supervised learning by providing ancillary training signals.

\paragraph{InterMPL-Last}
For SC-CTC,
InterMPL may not be an optimal choice,
as SC-CTC calculates intermediate losses using the same sequence targeted in the last layer.
Hence, we design another variant called InterMPL-Last.
Different from InterMPL (Fig.~\ref{fig:intermpl}(b) vs.\ Fig.~\ref{fig:intermpl}(c)),
InterMPL-Last utilizes only the final hypothesis of the offline model as pseudo-labels for calculating all the losses in the online model.
Given the $m$-th unlabeled sample $O_m\!\in\!\mathcal{D}_{\mathsf{unlab}}$ and
the last pseudo-labels generated by the offline model $W_m^{(K)}$,
the objective function of the online model is defined based on Eq.~\eqref{eq:L_ic} as
\begin{equation}
    \mathcal{L}_{\mathsf{ic}}(\hat{W}_m^{(K)} |O_m,\xi) = \sum_{k \in \mathcal{K}} \mathcal{L}_{\mathsf{ctc}}(\hat{W}_m^{(K)} | H^{(k)}, \xi).
\end{equation}
InterMPL-Last enables the online model to be trained on the most accurate pseudo-labels predicted by the offline model,
which permits more effective use of SC-CTC for semi-supervised training.

\vspace{-0.25cm}
\section{Experiments}
\vspace{-0.2cm}
\subsection{Experimental Setting}
\vspace{-0.15cm}
We used the ESPnet toolkit~\cite{watanabe2018espnet} for conducting the experiments, and
all the codes and recipes are made publicly available (see Sec.~\ref{sec:intro}).

\paragraph{Data}
We used LibriSpeech (LS)~\cite{panayotov2015librispeech} and TED-LIUM3 (TED3)~\cite{hernandez2018ted}.
LS is a corpus of read English speech,
containing 960 hours of training data (split into \textit{train-clean-100}, \textit{train-clean-360}, and \textit{train-other-500}).
TED3 is a corpus of English Ted Talks consisting of 450 hours of training data (\textit{train-ted3}).
We used the standard development and test sets of each dataset
for tuning hyper-parameters and evaluating performance, respectively.
As input speech features, 
we extracted 80 mel-scale filterbank coefficients with three-dimensional pitch features using Kaldi~\cite{povey2011kaldi}.
We used SentencePiece~\cite{kudo2018subword} to construct subword vocabularies from 
the \textit{train-clean-100} transcriptions.

\paragraph{Semi-supervised settings} 
We regarded \textit{train-clean-100} (LS-100) as the labeled data $\mathcal{D}_{\mathsf{lab}}$.
Based on a seed model trained on LS-100,
we simulated three semi-supervised settings using different unlabeled data $\mathcal{D}_\mathsf{unlab}$:
LS-100/LS-360, 
an in-domain setting using unlabeled \textit{train-clean-360} (LS-360); 
LS-100/LS-860, 
an in-domain setting using unlabeled \textit{train-\{clean-360,other-500\}} (LS-860); and 
LS-100/TED3, 
an out-of-domain setting using unlabeled \textit{train-ted3}. 

\paragraph{Model architecture}
We used the Conformer architecture~\cite{gulati2020conformer,guo2021recent}
consisting of two convolutional neural network layers followed by a stack of $18$ encoder blocks (i.e., $K\!=\!18$).
The number of heads, dimension of a self-attention layer, 
dimension of a feed-forward network, and kernel size were set to 
$4$, $256$, $1024$, and $7$, respectively.
Following~\cite{higuchi2022advancing},
we replaced batch normalization in the convolution module with group normalization with the group size of $4$.

\paragraph{Training and decoding configurations}
We trained the seed model for 150 epochs using the Adam optimizer~\cite{kingma2015adam} with $\beta_1\!=\!0.9$, $\beta_2\!=\!0.98$, $\epsilon\!=\!10^{-9}$, and
Noam learning rate scheduling~\cite{vaswani2017attention}.
We used 25k warmup steps and a learning rate factor of $5.0$.
The MPL training was iterated up to 200 epochs,
and the online model was trained using the Adam optimizer with an initial learning rate of $10^{-3}$, $\beta_1\!=\!0.9$, $\beta_2\!=\!0.999$, and $\epsilon\!=\!10^{-8}$.
The momentum coefficient $\alpha$, introduced in Sec.~\ref{ssec:mpl},
was decided following~\cite{higuchi2021momentum}.
The subword vocabulary size of CTC was set to $1024$.
SC-CTC and HC-CTC applied the intermediate loss to the 6th and 12th encoder layers (i.e., $\mathcal{K}\!=\!\{6,12,18\}$ in Eq.~\eqref{eq:L_ic}).
The output vocabulary size for each loss was set to $(1024, 1024, 1024)$ for SC-CTC and
$(256, 1024, 4096)$ for HC-CTC.
A final model was obtained for evaluation by averaging model parameters over ten checkpoints
that gave the best validation performance.
For the MPL-based methods,
we followed~\cite{higuchi2021momentum} and used the online model for evaluation.
During decoding,
we carried out the best path decoding of CTC~\cite{graves2006connectionist}.

\begin{table}[t]
    \centering
    \caption{WERs [\%] for models trained on fully labeled data. \texttt{A*}, \texttt{B*}, and \texttt{C*} indicate the oracle results for each semi-supervised setting. The LibriSpeech results are divided into ``test-\{clean / other\}'' sets.}
    \renewcommand{\arraystretch}{0.9}
    \label{tb:sup}
    \scalebox{.92}{
    \begin{tabular}{llcccc}
        \toprule
        & & LibriSpeech & TED-LIUM3 \\
        \cmidrule(l{0.3em}r{0.3em}){3-0} \cmidrule(l{0.3em}r{0.3em}){4-0}
        \textbf{Setting} & \textbf{Model} & \textbf{Test WER} ($\downarrow$) & \textbf{Test WER} ($\downarrow$) \\
        \midrule
        & \texttt{S1}\hspace{1mm} CTC & 8.4 / 23.1 & 26.7 \\
        LS-100 & \texttt{S2}\hspace{1mm} SC-CTC & 7.5 / 21.3 & 24.2 \\
        & \texttt{S3}\hspace{1mm} HC-CTC & \textbf{7.4} / \textbf{20.4} & \textbf{23.8} \\
        \midrule
        \multirow{3}{*}[-1pt]{\shortstack[l]{LS-100\\\ \ \ / LS-360}}
        & \texttt{A1}\hspace{1mm} CTC & 4.6 / 13.5 & -- \\
        & \texttt{A2}\hspace{1mm} SC-CTC & \textbf{3.9} / 12.0 & -- \\
        & \texttt{A3}\hspace{1mm} HC-CTC & 4.0 / \textbf{11.6} & -- \\
        \midrule
        \multirow{3}{*}[-1pt]{\shortstack[l]{LS-100\\\ \ \ / LS-860}}
        & \texttt{B1}\hspace{1mm} CTC & 3.5 / \pz8.8 & -- \\
        & \texttt{B2}\hspace{1mm} SC-CTC & \textbf{3.1} / \pz7.8 & -- \\
        & \texttt{B3}\hspace{1mm} HC-CTC & 3.2 / \textbf{\pz7.7} & -- \\
        \midrule
        \multirow{3}{*}[-1pt]{\shortstack[l]{LS-100\\\ \ \ / TED3}}
        & \texttt{C1}\hspace{1mm} CTC & -- & \pz7.5 \\
        & \texttt{C2}\hspace{1mm} SC-CTC & -- & \textbf{\pz6.8} \\
        & \texttt{C3}\hspace{1mm} HC-CTC & -- & \pz7.1 \\
        \bottomrule
    \end{tabular}}
\end{table}

\vspace{-0.2cm}
\subsection{Supervised Baseline and Oracle Results}
\vspace{-0.1cm}
\label{ssec:exp_oracle}
Table~\ref{tb:sup} shows the word error rate (WER) of seed models trained on LS-100 (\texttt{S*}) and
oracle models trained on fully labeled data in each semi-supervised setting (\texttt{A*}, \texttt{B*}, and \texttt{C*}).
Overall,
SC-CTC and HC-CTC outperformed CTC by relaxing the conditional independence assumption~\cite{nozaki2021relaxing,higuchi2022hierarchical}.
The quality of pseudo-labels is crucial for effective semi-supervised training, and
we can expect MPL to benefit from the seed models trained with the intermediate loss.

\begin{table}[t]
    \centering
    \caption{WER [\%] on in-domain LibriSpeech (LS) settings.}
    \label{tb:ls}
    \scalebox{.92}{
    \begin{tabular}{lllcc}
        \toprule
        & & & \multicolumn{2}{c}{\textbf{Test WER} ($\downarrow$)} \\
        \cmidrule(l{0.3em}r{0.3em}){4-5}
        \textbf{Setting} & \textbf{Method} & \textbf{Init.} & clean & other \\
        \midrule
        \multirow{4}{*}[-4pt]{\shortstack[l]{LS-100\\\ / LS-360}}
        & \texttt{X1}\hspace{1mm} MPL & \texttt{S1} (CTC) & 6.3 & 15.4 \\
        \cdashlinelr{2-5}
        & \texttt{X2}\hspace{1mm} InterMPL & \texttt{S2} (SC-CTC) & 5.7 & 14.5 \\
        & \texttt{X3}\hspace{1mm} InterMPL & \texttt{S3} (HC-CTC) & 5.5 & \textbf{14.1} \\
        & \texttt{X4}\hspace{1mm} InterMPL-Last & \texttt{S2} (SC-CTC) & \textbf{5.4} & \textbf{14.1} \\
        \midrule
        \multirow{4}{*}[-4pt]{\shortstack[l]{LS-100\\\ / LS-860}}
        & \texttt{Y1}\hspace{1mm} MPL & \texttt{S1} (CTC) & 6.0 & 11.9 \\
        \cdashlinelr{2-5}
        & \texttt{Y2}\hspace{1mm} InterMPL & \texttt{S2} (SC-CTC) & 5.4 & 11.0 \\
        & \texttt{Y3}\hspace{1mm} InterMPL & \texttt{S3} (HC-CTC) & 5.3 & \textbf{10.7} \\
        & \texttt{Y4}\hspace{1mm} InterMPL-Last & \texttt{S2} (SC-CTC) & \textbf{5.1} & \textbf{10.7} \\
        \bottomrule
    \end{tabular}}
    \vspace{-0.1cm}
\end{table}

\begin{figure}[t]
    \centering
    \includegraphics[width=0.94\columnwidth]{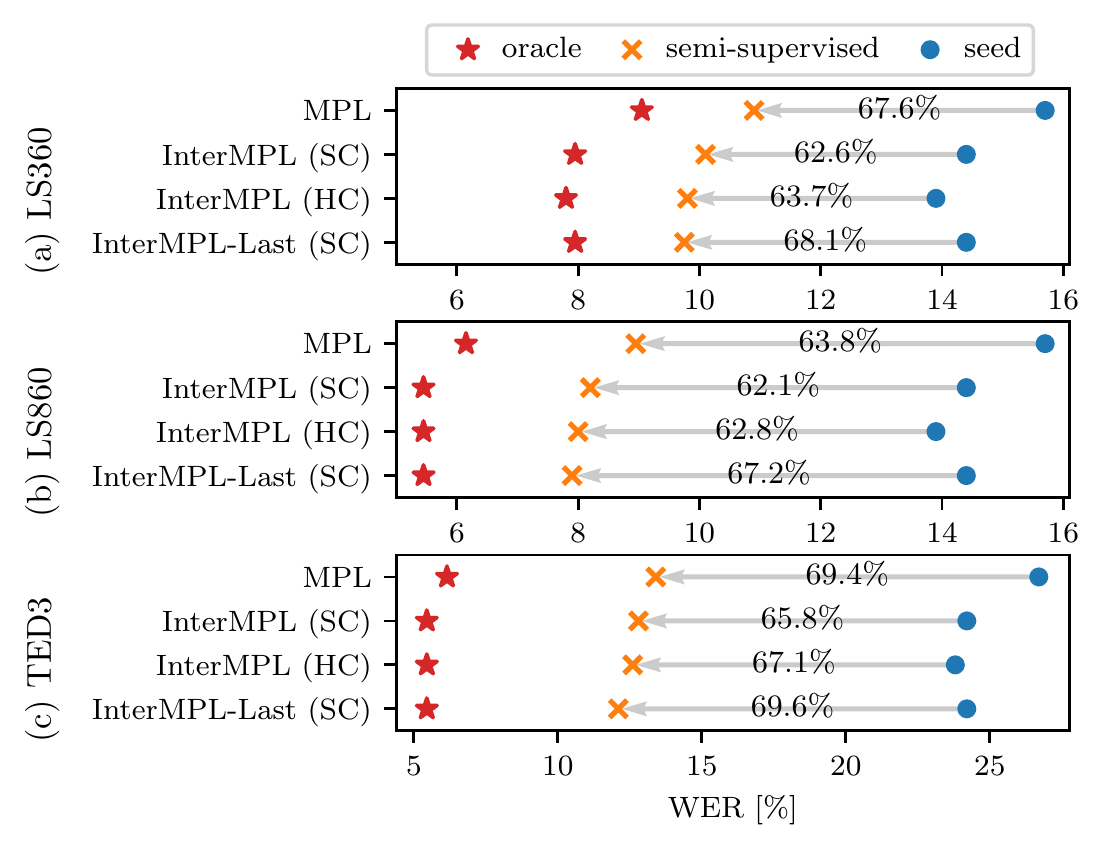}
    \vspace{-0.4cm}
    \caption{Visualization of WRR [\%] in each semi-supervised setting.}
    \vspace{-0.05cm}
    \label{fig:wrr}
\end{figure}

\begin{table}[t]
    \centering
    \caption{WER [\%] on out-of-domain LS-100/TED3.}
    \label{tb:ted3}
    \scalebox{.92}{
    \begin{tabular}{llc}
    \toprule
    \textbf{Method} & \textbf{Init.} & \textbf{Test WER} ($\downarrow$) \\
    \midrule
    \texttt{Z1}\hspace{1mm} MPL & \texttt{S1} (CTC) & 13.4 \\
    \cdashlinelr{1-3}
    \texttt{Z2}\hspace{1mm} InterMPL & \texttt{S2} (SC-CTC) & 12.8 \\
    \texttt{Z3}\hspace{1mm} InterMPL & \texttt{S3} (HC-CTC) & 12.6 \\
    \texttt{Z4}\hspace{1mm} InterMPL-Last & \texttt{S2} (SC-CTC) & \textbf{12.1} \\
    \bottomrule
    \end{tabular}}
    \vspace{-0.1cm}
\end{table}

\vspace{-0.05cm}
\subsection{Main Results}
\vspace{-0.1cm}
\paragraph{In-domain setting}
Table~\ref{tb:ls} shows WER on the LS settings,
comparing the conventional MPL~\cite{higuchi2022advancing} against the proposed InterMPL and InterMPL-Last.
In Fig.~\ref{fig:wrr}, we also compare the performance of each semi-supervised training method in the WER recovery rate (WRR)~\cite{ma2008unsupervised},
which shows how much performance gain is obtained
relative to the improvement from the seed to oracle WERs.
WRRs for LS are averaged on the clean and other sets.
Note that the seed models (\texttt{S*}) from Table~\ref{tb:sup} were used for the initialization in each method.
Looking at the results on the LS-360 setting (\texttt{X*}) in Table~\ref{tb:ls},
both InterMPL and InterMPL-Last led to distinct improvements over MPL (\texttt{X1} vs.\ \texttt{X2}, \texttt{X3}, \texttt{X4}),
indicating the effectiveness of using the well-trained seed models and applying intermediate CTC loss during semi-supervised training.
Comparing SC-CTC and HC-CTC-based InterMPL,
HC-CTC resulted in better performance by benefiting from using the pseudo-labels at different granularity (\texttt{X2} vs.\ \texttt{X3}).
InterMPL-Last was better suited for SC-CTC-based training than InterMPL (\texttt{X2} vs.\ \texttt{X4}),
as it was hypothesized that higher-quality labels are more appropriate for intermediate supervision.
Overall,
HC-CTC-based InterMPL and InterMPL-Last similarly achieved the best performance,
while InterMPL-Last gave higher WRRs in Fig.~\ref{fig:wrr}(a).

In the LS-860 setting with more unlabeled data (\texttt{Y*}) in Table~\ref{tb:ls},
the general trend was consistent with what was observed in the LS-360 setting.
In terms of WRR in Fig.~\ref{fig:wrr},
InterMPL-Last had the most significant gain, which was even higher than those of MPL.
Both InterMPL and InterMPL-Last were scalable to larger amounts of unlabeled data.

\paragraph{Out-of-domain setting}
Table~\ref{tb:ted3} lists results on the out-of-domain TED3 setting.
Both InterMPL and InterMPL-Last outperformed MPL (\texttt{Z1} vs.\ \texttt{Z2}, \texttt{Z3}, \texttt{Z4}),
demonstrating stable training on unlabeled data under the domain-mismatched condition.
In contrast to the in-domain results,
SC-CTC and HC-CTC-based InterMPL resulted in a similar performance (\texttt{Z2} vs.\ \texttt{Z3}), and
InterMPL-Last achieved lower WERs than InterMPL (\texttt{Z4} vs.\ \texttt{Z2}, \texttt{Z3}).
HC-CTC was less significant in the out-of-domain semi-supervised scenario,
including the oracle results in Table~\ref{tb:sup},
which we attribute to inferior generalization capability.
Subword vocabularies are constructed from the small LS-100 text set, and
the large vocabulary size used in HC-CTC (i.e., $4096$) was not generalized well to the TED3 domain.

\begin{table}[t]
    \centering
    \vspace{-0.2cm}
    \caption{Ablation study on LS-100/LS-360.}
    \centering
    \label{tb:ablation}
    \scalebox{.92}{
    \begin{tabular}{lccccc}
    \toprule
    & \multicolumn{2}{c}{\textbf{Test WER} ($\downarrow$)} & \multicolumn{2}{c}{\textbf{Test WRR} ($\uparrow$)} \\
    \cmidrule(l{0.3em}r{0.3em}){2-3} \cmidrule(l{0.3em}r{0.3em}){4-5}
    \textbf{Method} & clean & other & clean & other \\
    \midrule
    InterMPL (\texttt{X2}) & \textbf{\pz5.7} & \textbf{14.5} & \textbf{51.9} & \textbf{73.3} \\
    \ \ w/o inter. loss & \pz6.4 & 15.5 & 30.9 & 62.6 \\
    \cdashlinelr{1-5}
    InterMPL (\texttt{X3}) & \textbf{\pz5.5} & \textbf{14.1} & \textbf{55.3} & \textbf{72.0} \\
    \ \ w/o inter. loss & \pz5.8 & 14.8 & 45.8 & 63.8 \\
    \cdashlinelr{1-5}
    InterMPL-Last (\texttt{X4}) & \textbf{\pz5.4} & \textbf{14.1} & \textbf{59.2} & \textbf{77.0} \\
    \ \ w/ init. from \texttt{S1} & \pz5.9 & 14.4 & 46.4 & 74.1 \\
    \bottomrule
    \end{tabular}}
\end{table}

\vspace{-0.2cm}
\subsection{Ablation Study on Intermediate Loss}
\vspace{-0.1cm}
Table~\ref{tb:ablation} shows an ablation study validating the effectiveness of InterMPL.
We initialized a model using parameters of SC-CTC (\texttt{S2}) or HC-CTC (\texttt{S3}) and
performed standard MPL without intermediate loss.
Compared to the InterMPL results (\texttt{X2}, \texttt{X3}),
we observed that removing intermediate loss led to worsen WERs with degraded WRRs.
Interestingly,
MPL based on SC-CTC initialization resulted in a similar performance as that of vanilla MPL (\texttt{X1}).
This indicates the importance of applying intermediate loss during semi-supervised training.
We also performed InterMPL-Last initialized from CTC (\texttt{S1}),
which gave better results than those of MPL (\texttt{X1}).
The results suggest the importance of applying intermediate loss to both the seed model and semi-supervised training.

\vspace{-0.2cm}
\section{Conclusion}
\vspace{-0.2cm}
We proposed InterMPL,
a semi-supervised ASR method enhancing MPL using intermediate CTC loss.
We adopted SC-CTC or HC-CTC for training a seed model and
explored how pseudo-labels can be generated and used during semi-supervised training.
The experimental results and analysis revealed that InterMPL substantially outperforms MPL by fully using the intermediate loss mechanism.
Future work should explore using an external language model (LM) in InterMPL,
such as combining with LM-based PL~\cite{kahn2020self,higuchi2022advancing} and applying shallow fusion to intermediate predictions~\cite{komatsu2022better}.

\paragraph{Acknowledgement}
This work was supported in part by JST ACT-X (JPMJAX210J) and JSPS KAKENHI (JP21J23495).
This work was also based on results obtained from a project, JPNP20006, commissioned by the New Energy and Industrial Technology Development Organization (NEDO).

\newpage

\fontsize{8.4pt}{9.0pt}\selectfont
\bibliographystyle{IEEEtran}
\bibliography{refs}

\end{document}